\documentclass[aps,prd,a4paper,twocolumn,amsmath,showpacs,superscriptaddress,nofootinbib,preprintnumbers]{revtex4-1}
\pdfoutput=1
\usepackage{amssymb,amsmath,latexsym,mathrsfs}
\usepackage[sort&compress]{natbib}
\usepackage{graphicx,subfigure}
\usepackage{epsfig}
\usepackage{varioref,xr-hyper}
\usepackage{color}
\usepackage{multirow}
\usepackage{array}
\usepackage{hyperref}
\usepackage{wasysym}
\usepackage{color}
\usepackage{float}
\usepackage{xcolor}
\usepackage[utf8]{inputenc}
\usepackage[T1]{fontenc}

\newcommand{\be}{\begin{equation}}
\newcommand{\ee}{\end{equation}}
\newcommand{\bea}{\begin{eqnarray}}
\newcommand{\eea}{\end{eqnarray}}
\newcommand{\mc}{\mathcal}
\newcommand{\vo}{\mathcal{V}}

\begin{document}

\title{A Fake Instability in String Inflation}

\author{Michele Cicoli}
\email{michele.cicoli@unibo.it}
\affiliation{Dipartimento di Fisica e Astronomia, Universit\'a di Bologna, via Irnerio 46, 40126 Bologna, Italy}
\affiliation{INFN, Sezione di Bologna, via Irnerio 46, 40126 Bologna, Italy}

\author{Veronica Guidetti}
\email{veronica.guidetti@desy.de}
\affiliation{Deutsches Elektronen-Synchrotron (DESY), 22607 Hamburg, Germany}

\author{Francesco Muia,}
\email{fm538@damtp.cam.ac.uk}
\affiliation{DAMTP, University of Cambridge, Wilberforce Road, Cambridge, CB3 0WA, UK}

\author{Francisco G. Pedro}
\email{francisco.soares@unibo.it}
\affiliation{Dipartimento di Fisica e Astronomia, Universit\'a di Bologna, via Irnerio 46, 40126 Bologna, Italy}
\affiliation{INFN, Sezione di Bologna, via Irnerio 46, 40126 Bologna, Italy}

\author{Gian Paolo Vacca}
\email{vacca@bo.infn.it}
\affiliation{INFN, Sezione di Bologna, via Irnerio 46, 40126 Bologna, Italy}

\begin{abstract}
In type IIB Fibre Inflation models the inflaton is a K\"ahler modulus which is kinetically coupled to the corresponding axion. In this setup the curvature of the field space induces tachyonic isocurvature perturbations normal to the background inflationary trajectory. However we argue that the associated instability is unphysical since it is due to the use of ill-defined entropy variables. In fact, upon using the correct relative entropy perturbation, we show that in Fibre Inflation axionic isocurvature perturbations decay during inflation and the dynamics is essentially single-field.
\end{abstract}

\maketitle

\section{Introduction}

A class of well-studied string inflationary models is Fibre Inflation (FI) whose name originates from the fact the inflaton is a type IIB K\"ahler modulus controlling the size of a K3 or $T^4$ fibre over a $\mathbb{P}^1$ base. These models have been built with the framework of Large Volume Scenarios \cite{Balasubramanian:2005zx, Cicoli:2008va}. The inflaton is a leading order flat direction whose potential can be generated by different combinations of perturbative corrections: 1-loop open string Kaluza-Klein (KK) and winding effects \cite{Cicoli:2008gp}, 1-loop KK corrections and higher order $\alpha'$ terms \cite{Broy:2015zba}, or 1-loop winding contributions and $\alpha'$ corrections \cite{Cicoli:2016chb}. 

Besides moduli stabilisation, these models features several promising properties, including an approximate shift symmetry for the inflaton potential \cite{Burgess:2016owb, Burgess:2014tja}, global Calabi-Yau orientifold constructions with chiral matter \cite{Cicoli:2011it, Cicoli:2016xae, Cicoli:2017axo}, and a detailed understanding of the post-inflationary evolution. In particular, preheating effects turn out to be negligible \cite{Antusch:2017flz} while standard perturbative reheating \cite{Cabella:2017zsa, Cicoli:2018cgu} can lead to an epoch of radiation domination with initial temperature which is low enough to avoid any decompactification due to finite-temperature effects \cite{Anguelova:2009ht}. Together with Standard Model particles, the inflaton decay produces also ultra-light bulk axions behaving as extra relativistic species which contribute to $N_{\rm eff}$. 

Interestingly, the potential of FI models resembles Starobinsky inflation \cite{Starobinsky:1980te} and supergravity $\alpha$-attractors \cite{Kallosh:2013maa,Kallosh:2017wku} since it features a trans-Planckian plateau followed by a steepening at large inflaton values that can produce a CMB power loss at large scales \cite{Cicoli:2013oba, Pedro:2013pba, Cicoli:2014bja} and primordial black hole dark matter \cite{Cicoli:2018asa}. Moreover the extra-dimensional geometry constrains the inflaton field range to values of $\mc{O}(5)$ in Planck units \cite{Cicoli:2018tcq}. This, in turn, translates in a tensor-to-scalar ratio $r \lesssim 0.01$. A recent work \cite{Cicoli:2020bao} determined the values of the microscopic parameters of FI models which give the best fit to most recent cosmological data, finding at $68\%$ CL $n_s = 0.9696^{+0.0010}_{-0.0026}$, $r = 0.00731^{+0.00026}_{-0.00072}$ and $N_{\rm eff} = 3.062^{+0.004}_{-0.015}$ (for Planck 2018 temperature and polarisation data only). 

Despite all these interesting features, it has been recently pointed out \cite{Cicoli:2018ccr, Cicoli:2019ulk} that FI models might be plagued by a geometrical instability \cite{Gong:2011uw,Renaux-Petel:2015mga}. More precisely, isocurvature perturbations associated to one of the two ultra-light axions typical of FI models, experience a growth during inflation triggered by the curvature of the underlying field space. At first sight, this effect might seem dangerous since it would bring the system away from the perturbative regime. However, as already noticed in \cite{Cicoli:2019ulk}, the background trajectory remains stable. 

In this paper we shall resolve this paradox by exploiting the analysis performed in \cite{Cicoli:2021yhb} that clarified which is the correct entropy variable that should be used to match the evolution of the isocurvature modes between inflation and radiation dominance. In fact, we shall show that the geometrical instability of FI models is just apparent since it is an artifact of the decomposition of a generic perturbation into modes tangent and orthogonal to the inflationary trajectory. The spurious nature of the instability resides in the fact that the normal unit vector diverges, while no tachyonic mass for the cosmological perturbations is seen when using the original field basis. According to the analysis performed in \cite{Cicoli:2021yhb}, we therefore used the proper variable, the relative entropy perturbation, which is both gauge invariant and finite, and found that isocurvature perturbations indeed decay during inflation, in full agreement with the fact that the background dynamics is stable and essentially single-field.

We therefore conclude that FI models are not plagued by any geometrical destabilisation effect, satisfy current isocurvature perturbation bounds, and the inflationary evolution of the system remains always in the regime where perturbation theory works very well. 

This paper is organised as follows. In Sec. \ref{FibreReview} we first review the main features of FI models and the origin of the apparent destabilisation effect. In Sec. \ref{NoInst} we then show the absence of any instability by studying the evolution of the relative entropy perturbation. We finally present our conclusions in Sec. \ref{Conclusions}.

\section{A geometrical instability in Fibre Inflation?}
\label{FibreReview}

\subsection{Basics of Fibre Inflation}

All FI models are qualitatively very similar, and so, without loss of generality, we will focus on the original formulation \cite{Cicoli:2008gp} which involves type IIB Calabi-Yau orientifold compactifications with fluxes, D3/D7-branes, O3/O7-planes and $h^{1,1}=3$ K\"ahler moduli $T_i=\tau_i +{\rm i} \theta_i$, $i=1,2,3$. The internal volume looks like:
\be
\vo=\alpha(\tau_b \sqrt{\tau_f}-\lambda_s \tau_s^{3/2})\,,
\ee
where $\alpha$ and $\lambda_s$ are $\mc{O}(1)$ constants (which depend on the intersection numbers), $\tau_s$ is a blow-up mode, $\tau_b$ is the base modulus and $\tau_f$ controls the volume of the fibre K3 or $T^4$ divisor. In a large volume expansion, the moduli potential receives contributions at 3 different orders: ($i$) at leading order, $T_s$-dependent non-perturbative effects and $\mc{O}(\alpha'^3)$ corrections stabilise $\vo$, $\tau_s$ and $\theta_s$ at $\vo\gg 1$ giving them a mass larger than the Hubble scale during inflation; ($ii$) at subleading order KK and winding 1-loop open string effects develop the inflationary potential for $\tau_f$; ($iii$) the two axions $\theta_b$ and $\theta_f$ are almost massless and much lighter than $\tau_f$ since they become massive only via highly suppressed $\vo$-dependent non-perturbative effects. 

The inflationary potential in terms of the canonically normalised inflaton $\phi$ reads (setting the reduced Planck mass $M_p = 1$):
\be
V_{\rm inf} = V_0\left[ 3-4 \,e^{-\frac{\phi}{\sqrt{3}} }  +  e^{-\frac{4\phi}{\sqrt{3}} }+ R\left(e^{\frac{2\phi}{\sqrt{3}} } - 1\right)\right],
\label{Vinf}
\ee
with:
\be
V_0 = \frac{g_s^{1/3} W_0^2 A}{8\pi \lambda^2 \vo^{10/3}}\qquad\text{and}\qquad
R = 16 g_s^4 \frac{AC}{B^2}\,,
\ee
where, following the notation of \cite{Cicoli:2018cgu}, $g_s$ is the string coupling, $W_0$ is the flux-generated $\mc{O}(10-100)$ superpotential, $A$, $B$ and $C$ are $\mc{O}(1)$ flux-dependent coefficients of the string loop corrections, and $\lambda = (4A/B)^{2/3}$. The best fit analysis of \cite{Cicoli:2020bao} found $R<4.80\times 10^{-6}$ and $10^{11}\, V_0=6.76^{+0.25}_{-0.49}$ for Planck data alone at $68\%$ CL. Given that for $R<4.80\times 10^{-6}$, horizon exit occurs always in the plateau region where the term proportional to $R$ is negligible, in what follows we shall simply set $R=0$ (which would imply no power loss at large angular scales). In this case the relation between the scalar spectral index $n_s$ and the tensor-to-scalar ratio $r$ can approximated as $r\simeq 6 (n_s-1)^2$ which reproduces rather well the best fit values $n_s=0.9696^{+0.0010}_{-0.0026}$ and $r=0.00731^{+0.00026}_{-0.00072}$ found in \cite{Cicoli:2020bao}. Notice that such a large value of $r$ could be tested by the next generation of cosmological observations.

The reheating temperature from the inflaton decay $T_{\rm rh}$ can be written as $T_{\rm rh} = 3\,\gamma\cdot 10^{10}$ GeV where $\gamma = 2 \lambda \alpha_{\rm vis}  g_s^{4/3} \vo^{2/3}$ (with $\alpha_{\rm vis} = g^2/(4\pi)$) controls the branching ratio for the inflaton decay into visible sector gauge bosons and the ultra-light axions $\theta_b$ and $\theta_f$ which yield extra relativistic degrees of freedom parametrised by $\Delta N_{\rm eff}$. Given that the number of efoldings of inflation $N$ depends on $T_{\rm rh}$, $\gamma$ determines both $N$ and $\Delta N_{\rm eff}$ as \cite{Cicoli:2018cgu}:
\be
N =  52+\frac13 \ln\gamma 
\qquad\text{and}\qquad
\Delta N_{\rm eff} = \frac{0.6}{\gamma^2}\,, 
\label{NeffPred}
\ee
where the best fit for Planck data alone is $7.41<\gamma\lesssim 20$ (which implies $N\simeq 52$) and $N_{\rm eff}=3.062^{+0.004}_{-0.015}$ at $68\%$ CL \cite{Cicoli:2020bao}. It is straightforward to check that all these observational constraints, combined with the requirement of having an effective field theory under control, can be satisfied for rather natural choices of the underlying parameters $W_0$, $A$, $B$ and $C$, together with $0.065\lesssim g_s\lesssim 0.125$ and $2500\lesssim \vo \lesssim 9000$.

\subsection{Unstable isocurvature modes?}

\begin{figure}[!htbp]
\centering
\includegraphics[scale = 0.5]{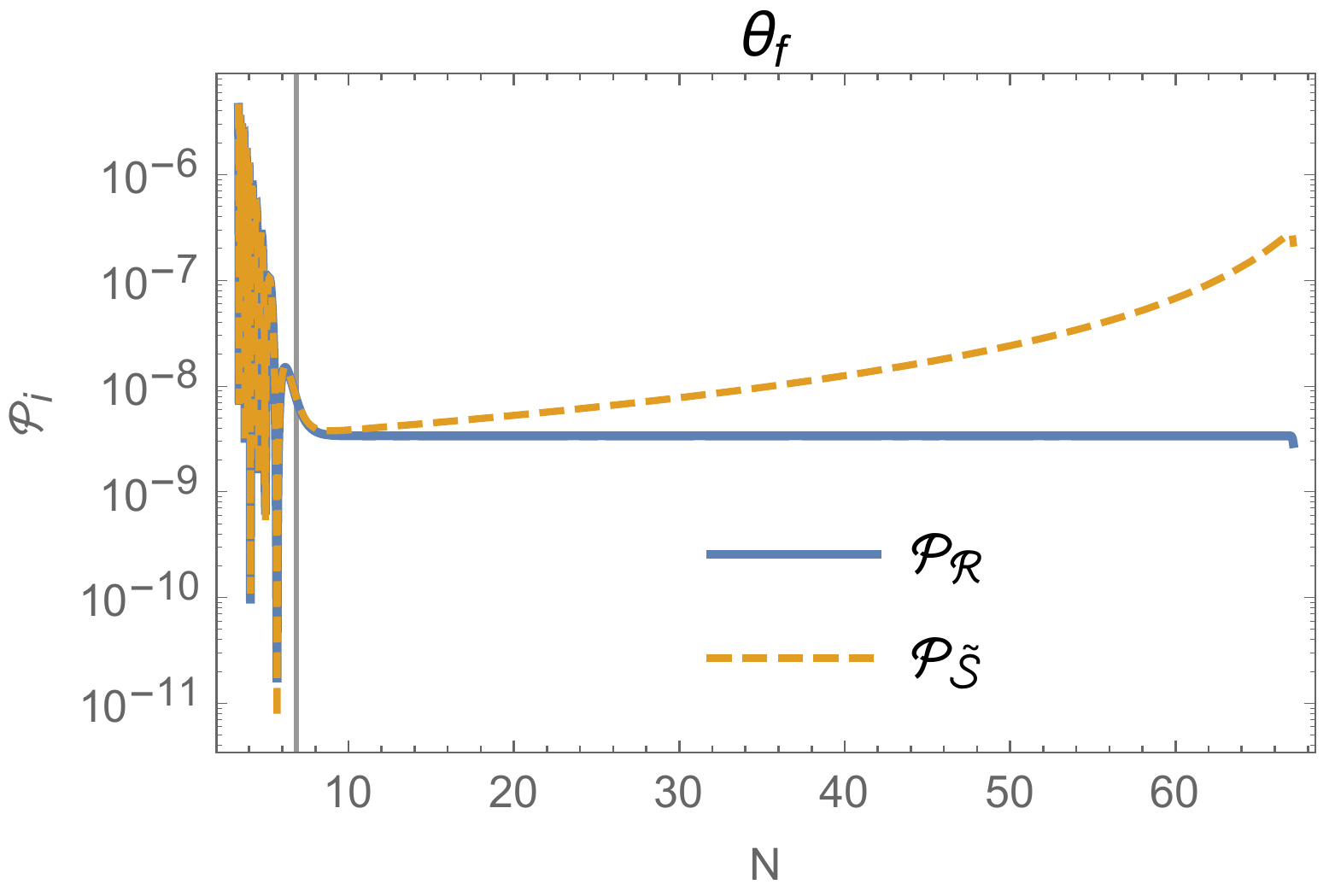}
\includegraphics[scale = 0.5]{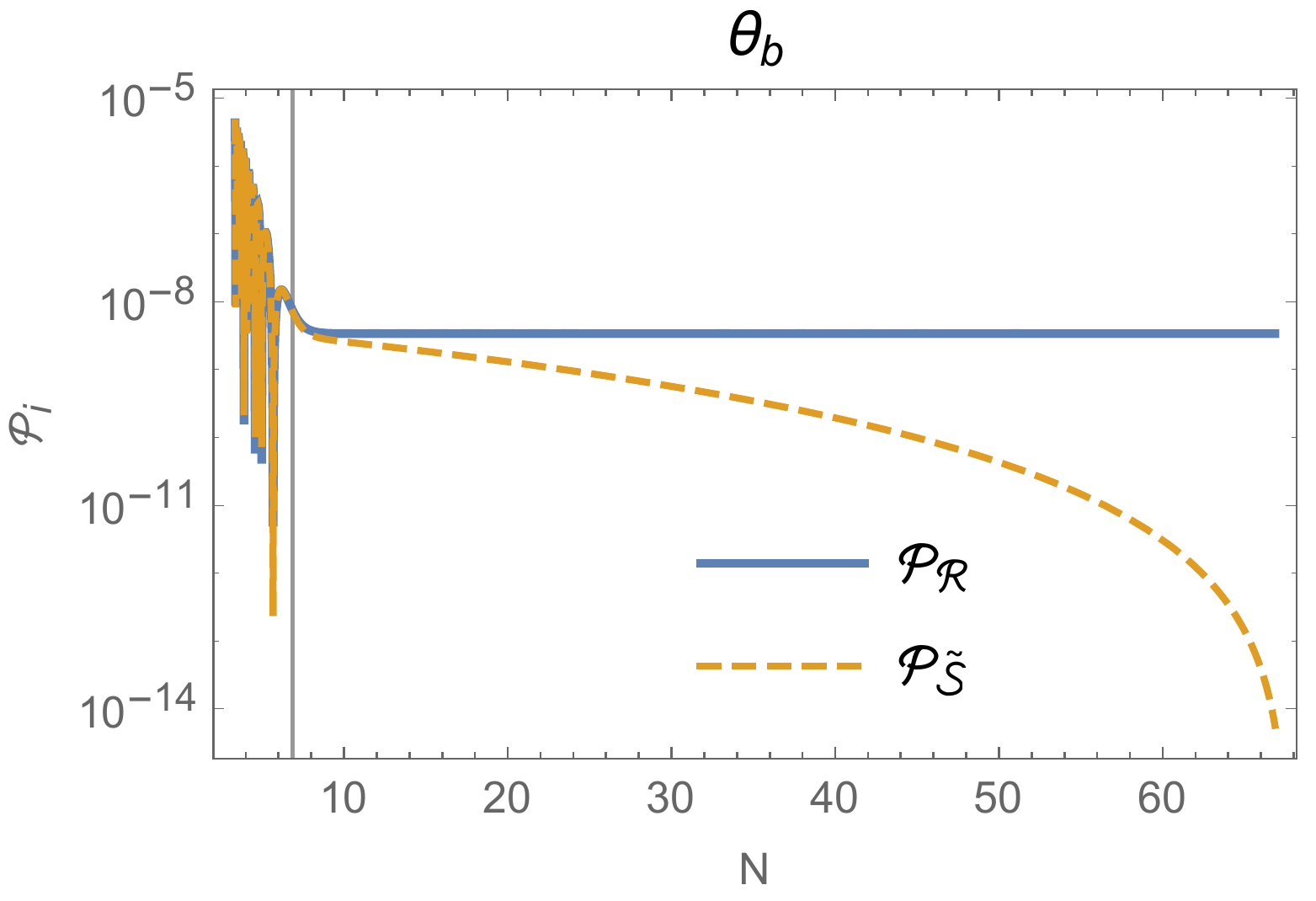}
\caption{Super-horizon evolution of curvature ($\mc{P}_{\mc{R}}$) and isocurvature ($\mc{P}_{\tilde{\mc{S}}}$) power spectra (in terms of $\tilde{\mc{S}}\equiv \delta s/\sqrt{2\epsilon}$) for modes exiting the horizon $60$ e-foldings before the end of inflation. While $\theta_b$ entropy modes decay, the isocurvature perturbations associated to $\theta_f$ experience an exponential growth.}
\label{Fig1}
\end{figure}

The fields $\vo$, $\tau_s$ and $\theta_s$ are heavier than the inflaton during inflation, and so remain fixed at their minima. On the other hand, the two ultra-light axions $\theta_b$ and $\theta_f$ source isocurvature perturbations. These axionic fields are kinetically coupled to the inflaton since the Lagrangian contains terms of the form \cite{Cicoli:2018ccr}:
\be
\mc{L}\supset \frac12 h^2(\phi)\,\partial_\mu \theta_b\partial^\mu \theta_b + \frac12  f^2(\phi)\,\partial_\mu \theta_f \partial^\mu \theta_f\,,
\ee
where:
\be
h(\phi) = \frac{\alpha\sqrt{c}}{\vo^{2/3}}\,e^{\frac{1}{\sqrt{3}}\phi}
\qquad\text{and}\qquad
f(\phi) = \frac{e^{-\frac{2}{\sqrt{3}}\phi}}{\sqrt{2} c \vo^{2/3}}\,,
\label{KinCoupl}
\ee
with $c = g_s^{4/3} \lambda$. These kinetic couplings correspond to a curved field space which induces a tachyonic entropy perturbation $\delta s$ associated to $\theta_f$ \cite{Cicoli:2018ccr}, as we now briefly review. The entropy perturbation variable $\delta s$ has been introduced in \cite{Gordon:2000hv} and corresponds to perturbations orthogonal to the background inflationary trajectory. Considering the 2D ($\phi,\theta_f$) subspace obtained by keeping the axion $\theta_b$ fixed at its minimum, $\delta s$ is defined as:
\be
\delta s = N_\phi\, \delta\phi + N_f\, \delta\theta_f\,,
\ee
where $N_\phi$ and $N_f$ are the components of the normal unit vector $\vec{N}$ given by \cite{Achucarro:2010da,Cremonini:2010sv}:
\be
\vec{N} = \left(\begin{matrix}
           N_\phi \\
           N_f \\
           \end{matrix}\right)
           = \frac{f}{\sqrt{\dot\phi^2 + (f \dot\theta_f)^2}}\left(\begin{matrix}
           - \dot\theta_f \\
           \dot\phi \\
           \end{matrix}\right).
\ee
When $\theta_b$ is massless, the effective mass-squared of $\delta s$ evaluated on the background attractor trajectory receives contributions from the metric connection and the Ricci scalar of the field manifold $R=-8/3$ which give \cite{Cicoli:2019ulk}:
\be
m_{\delta s}^2\simeq -\frac{2}{\sqrt{3}}\left( \partial_\phi V_{\rm inf} + \frac{2}{\sqrt{3}}\,\dot{\phi}^2\right)<0 \,.
\label{eq:meffULF}
\ee
This mass is clearly tachyonic since $\partial_\phi V_{\rm inf}>0$. Notice that the isocurvature perturbation associated to $\theta_b$ would have instead a positive mass-squared due to the different sign in the exponent of the corresponding kinetic coupling, as can be seen from (\ref{KinCoupl}). The geometrical instability associated to $\theta_f$ might seem very dangerous since it induces an exponential growth of isocurvature modes that could bring the system away from the perturbative approximation, as can be seen from Fig. \ref{Fig1}. 

This consideration appears however to be in contradiction with the analysis of \cite{Cicoli:2018ccr} which showed that, regardless of the choice of initial conditions, the system quickly approaches a single field dynamics where the inflationary trajectory has a vanishing turning rate and the ultra-light axions are frozen with zero velocity. In Sec. \ref{NoInst} we shall shed light on this paradox, arguing that in this case $\delta s$ is an ill-defined variable since the exponential growth is hidden in the $N_f$ component of the normal vector.

\begin{figure}[!htbp]
\centering
\includegraphics[scale = 0.5]{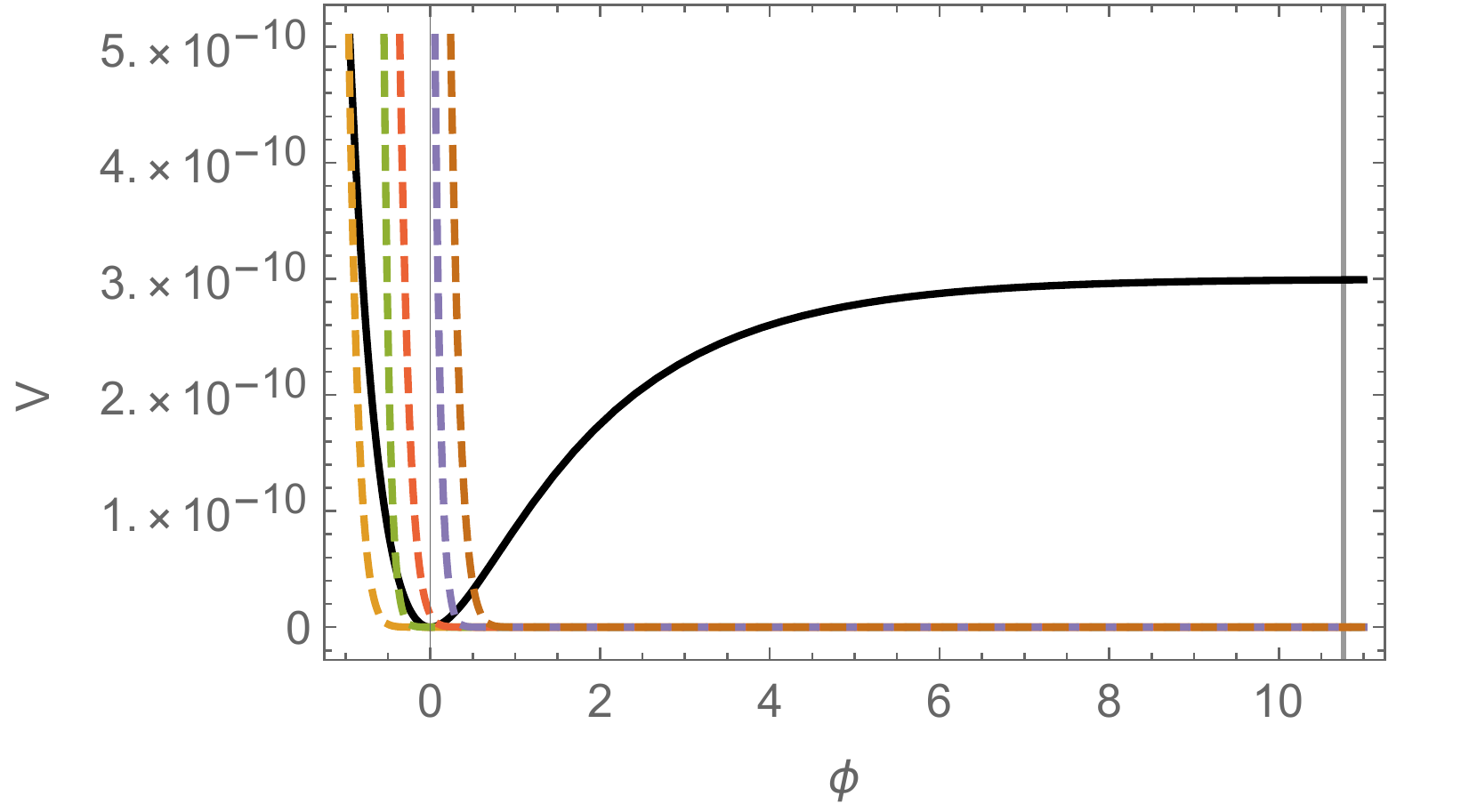}
\caption{Hierarchy between the inflationary potential (\ref{Vinf}) with $V_0=2\times 10^{-10}$ and $R=0$ (denoted by the black line) and the potential for $\theta_f$ (\ref{Vf}) with $c=0.04$ which gives $\langle\tau_f\rangle=4$ (corresponding to $\phi=0$) and $\langle\tau_b\rangle=500$ (for $\alpha=1$). The vertical line denotes the maximum inflaton value compatible with a controlled effective field theory which we choose to be $\tau_b>\tau_{b,{\rm min}} = 1$ corresponding to $\phi <\phi_{\rm max} = \sqrt{3} \ln\langle\tau_b\rangle \simeq 10.8$. The dashed lines correspond to different values of $W_0$ and $n$ as follows: $n=1$ and $W_0=0.25$ (yellow); $n=1$ and $W_0=25$ (green); $n=2$ and $W_0=0.25$ (red); $n=2$ and $W_0=25$ (violet); $n=4$ and $W_0=0.25$ (brown).}
\label{Fig2}
\end{figure}

Before showing that FI models are stable, let us stress that $m_{\delta s}$ would be tachyonic also when considering a massive $\theta_f$ axion. The scalar potential for $\theta_f$ is generated by non-perturbative effects and looks like:
\be 
V_f (\phi,\theta_f)= \Lambda \,e^{\frac{2}{\sqrt{3}}\phi} e^{-k\, e^{\frac{2}{\sqrt{3}}\phi}}\cos{\left(\frac{2\pi}{n}\theta_f\right)}\,,
\label{Vf}
\ee
where (setting the prefactor of non-perturbative effects to unity):
\be
\Lambda = \frac{8\pi c W_0}{n \vo^{4/3}} \qquad\text{and}\qquad k = \frac{2\pi c}{n}\, \vo^{2/3}\,.
\ee
The total potential now becomes:
\be
V_{\rm tot} = V_{\rm inf}(\phi) + V_f (\phi, \theta_f)\,,
\ee
where $V_{\rm inf}$ is given by (\ref{Vinf}) and we need to require $V_f (\phi,\theta_f)\ll V_{\rm inf}(\phi)$ to prevent any modification of the FI dynamics. Due to the double exponential suppression in (\ref{Vf}), this condition implies that $V_f$ can make $m_{\delta s}^2$ positive only locally in field space but not throughout the whole trans-Planckian inflaton range, $\Delta\phi\simeq \mc{O}(5)$. In fact, \cite{Cicoli:2019ulk} has shown that $m_{\delta s}^2<0$ for any choice of the microscopic parameters which keeps $V_f (\phi,\theta_f)\ll V_{\rm inf}(\phi)$ for the whole inflationary epoch. Fig. \ref{Fig2} shows how the hierarchy between $V_f (\phi,\theta_f)$ and $ V_{\rm inf}(\phi)$ varies as a function of $\Lambda$ and $n$. 

\section{Stability of Fibre Inflation and decaying isocurvature modes}
\label{NoInst}

\subsection{Physical entropy variable}

The quantity constrained by Planck observations is the primordial isocurvature fraction $\beta_{\rm iso} = \mc{P}_{\mc{S}_{i\gamma}}/ [\mc{P}_{\mc{R}} + \mc{P}_{\mc{S}_{i\gamma}}]$, where $\mc{P}_{\mc{R}}$ is the curvature power spectrum while $\mc{S}_{i\gamma}$ is the relative entropy perturbation between photons and a different $i$-th species (cold dark matter, baryons or neutrinos), with $\beta_{\rm iso}<\mc{O}(0.1-0.01)$ depending on the species involved \cite{Akrami:2018odb}. Thus in order to compare the predictions of FI models with observations, one would need to focus on the super-horizon evolution of the relative entropy perturbation between $\phi$ and $\theta_f$ defined as \cite{Malik:2004tf}:
\be 
\mc{S}_{\phi \theta_f}= -3 H \left(\frac{\delta\rho_\phi}{\dot\rho_\phi}-\frac{\delta\rho_{\theta_f}}{\dot\rho_{\theta_f}}\right),
\ee
where $\rho_i$ are the energy densities of the two fields. Thanks to the detailed analysis of reheating performed in \cite{Cicoli:2018cgu}, one should then derive $\mc{S}_{i\gamma}$ from $\mc{S}_{\phi \theta_f}$. However, as pointed out in \cite{Cicoli:2021yhb} (see also \cite{Wands:2000dp}), in FI models $\mc{S}_{\phi \theta_f}$ is an ill-defined quantity (despite being gauge invariant even for a curved field space) since $\dot\rho_{\theta_f}\to 0$ given that the energy density of the ultra-light axions vanishes. Thus, similarly to $\delta s$, also $\mc{S}_{\phi \theta_f}$ would yield an unphysical divergence of isocurvature perturbations.

As explained in \cite{Cicoli:2021yhb}, the correct physical, i.e. both gauge invariant and finite, entropy variable which should be used in this case is the relative entropy perturbation $\mc{S}_{\rm rel}$ which can be defined starting from the notion of total entropy perturbation $\mc{S}$:
\be
\mc{S}=\frac{H}{\dot{P}}\,\delta P_{\rm nad}\,,
\label{eq:S}
\ee
where $\delta P_{\rm nad}$ is the non-adiabatic pressure perturbation which can be obtained from the total pressure perturbation $\delta P$ as $\delta P_{\rm nad}=\delta P-c_s^2 \delta\rho$ with $c_s^2=\dot P/\dot \rho$. 

The relative entropy perturbation $\mc{S}_{\rm rel}$ is then obtained by subtracting the intrinsic entropy perturbation $\mc{S}_{\rm int}$ from the total one, $\mc{S}_{\rm rel}= \mc{S}-\mc{S}_{\rm int}$, where $\mc{S}_{\rm int}$ is given by the sum of the entropies associated to each fluid $\mc{S}_{{\rm int},i}$ \cite{Malik:2004tf}:
\be
\mc{S}_{\rm int}=\sum_i \mc{S}_{{\rm int},i}=\sum_i\frac{H}{\dot{P}}\left(\delta P_i - c_i^2 \delta \rho_i\right),
\label{eq:Sint}
\ee
with $c_i^2=\dot P_i/\dot\rho_i$ denoting the sound speed of each scalar cosmological fluid. Using $\delta P = \sum_i \delta P_i$ and $\delta \rho = \sum_i \delta \rho_i$, the relative entropy perturbation hence becomes (focusing on the FI case with two fields, $\phi$ and $\theta_f$):
\be
\mc{S}_{\rm rel} = \frac{\left(c_{\theta_f}^2-c_\phi^2\right)}{3\dot\rho\dot P}\, \dot\rho_\phi \dot\rho_{\theta_f}  \mc{S}_{\phi \theta_f}\,.
\label{Srel}
\ee
This quantity is now well-behaved since its denominator is independent on the vanishing quantity $\dot\rho_{\theta_f}$. The prescription of \cite{Cicoli:2021yhb} is to study the evolution of $\mc{S}_{\rm rel}$ from inflation to radiation dominance after reheating, and then to infer from (\ref{Srel}) $\mc{S}_{i\gamma}$ and the final prediction for $\beta_{\rm iso}$. 

\subsection{Decaying isocurvature perturbations}

We shall now focus on FI models and show that the power spectrum of isocurvature modes associated to $\mc{S}_{\rm rel}$ decays on super-horizon scales during inflation. We start by rewriting (\ref{Srel}) in a form which is easier to evaluate analytically: 
\be
\mc{S}_{\rm rel}=\frac{H}{\dot P} \left[\left( c_\phi^2-c_s^2\right)\delta\rho_\phi+\left( c_{\theta_f}^2-c_s^2\right)\delta\rho_{\theta_f}\right]\,.
\label{Srelnew}
\ee
The energy and pressure of the two fields can be written as: 
\bea
\rho_\phi&=&\frac12 \dot\phi^2+V_{\rm inf}\,,\qquad \rho_{\theta_f}=\frac12 (f\dot\theta_f)^2+V_f\,, \nonumber \\
P_\phi&=&\frac12 \dot\phi^2-V_{\rm inf}\,,\qquad P_{\theta_f}=\frac12 (f\dot\theta_f)^2-V_f\,,
\eea
where this split does not have a clear physical meaning since $V_f(\phi,\theta_f)$. It is however useful to evaluate $\mc{S}_{\rm rel}$ and to reduce to a single field dynamics since we will see that for $V_f(\phi,\theta_f)\ll V_{\rm inf}(\phi)$ the system very quickly approaches an attractor background trajectory characterised by $(f\dot\theta_f)\to 0$ and $\rho_{\theta_f}\to 0$.

The total sound speed of the system is given by:
\be
c_s^2=1+\frac{2}{3H}\frac{\dot\phi\,\partial_\phi V_{\rm tot}  + \dot\theta_f\,\partial_{\theta_f} V_f }{\dot\phi^2 +(f\dot\theta_f)^2}\,,
\ee
while the sound speeds of the fluid components are:
\bea
c_\phi^2&=&1+\frac{2\,\partial_\phi V_{\rm inf}}{3H\dot\phi-f \partial_\phi f\, \dot\theta_f^2 + \partial_\phi V_f}\,, \nonumber \\
c_{\theta_f}^2&=&1+\frac{2\left( \partial_{\theta_f} V_f \dot\theta_f+ \partial_\phi V_f\dot\phi\right)}{\dot\theta_f^2 \left( 3H f^2+ f \partial_\phi f \dot\phi -\partial_\phi V_f \dot\phi\dot\theta_f^{-2}\right)}\,,
\eea
where we used the equations of motion given by:
\bea
\ddot\phi &=& -3H\dot\phi +f\partial_\phi f \dot\theta_f^2-\partial_\phi V_{\rm tot}\,,  \\
f^2\ddot\theta_f &=& -3H f^2\dot\theta_f-2 f\partial_\phi f \dot\phi\dot\theta_f-\partial_{\theta_f} V_f\,.
\label{eq2}
\eea
To compute the energy density fluctuations, we use perturbation theory at linear order in the spatially flat gauge (since $\mc{S}_{\rm rel}$ is gauge invariant), obtaining:
\bea
\delta\rho_\phi&=&-\Phi \dot\phi^2 + \dot\phi \delta\dot\phi  +\partial_\phi V_{\rm inf} \delta\phi\,,  \\
\delta\rho_{\theta_f}&=& (f \dot\theta_f)^2 \left(\frac{\partial_\phi f}{f}\delta\phi +\frac{\delta\dot\theta_f}{\dot\theta_f}-\Phi \right)+ \partial_\phi V_f \delta\phi+\partial_{\theta_f} V_f \delta\theta_f \,, \nonumber
\eea
where the lapse function $\Phi$ reads:
\be 
\Phi=\frac{1}{2H}\left(\dot\phi \delta\phi + f^2  \dot\theta_f \delta\theta_f\right)\,.
\ee
In order to obtain simple analytic results, we now consider the case with $V_f=0$ which gives a very good approximation of the generic behaviour of the system since $V_f\ll V_{\rm inf}$ (we have however performed a numerical analysis also for $V_f\neq 0$ whose results we present below). 

The quantities derived above have to be evaluated on the inflationary trajectory. As derived in \cite{Cicoli:2018ccr}, the equation of motion (\ref{eq2}) admits a slow-roll solution which looks like (prime denotes a derivative with respect to $N$):
\be
(f\theta_f')\simeq (f\theta_f')(0)\,e^{-3N} \to 0\,, 
\ee
which implies that during inflation the ultra-light axion $\theta_f$ very quickly gets frozen. It is easy to realise that in this limit $(c_\phi^2 - c_s^2)\to 0$ and $\delta\rho_{\theta_f}\to 0$, while both $(c_{\theta_f}^2-c_s^2)$ and $\delta\rho_\phi$ remain finite (in particular $c_{\theta_f}^2\to 1$). Moreover $H/\dot P\to -1/\left(6\epsilon H^2 + 2 \partial_\phi V \sqrt{2\epsilon} \right)$ , and so also this ratio is finite. Thus both terms in (\ref{Srelnew}) vanish and $\mc{S}_{\rm rel}\to 0$ in the attractor inflationary trajectory.

Isocurvature modes associated to the relative entropy perturbation $\mc{S}_{\rm rel}$ are therefore negligible in FI models. The geometrical destabilisation found in \cite{Cicoli:2018ccr} and reviewed in Sec. \ref{FibreReview} is thus a spurious effect due to the use of an ill-defined entropy variable. We conclude that the dynamics of FI models is stable and essentially single-field, begin characterised by decaying isocurvature modes which give rise to a negligibly small $\beta_{\rm iso}$ in full agreement with Planck data. 

\begin{figure}[!htbp]
\centering
\includegraphics[scale = 0.5]{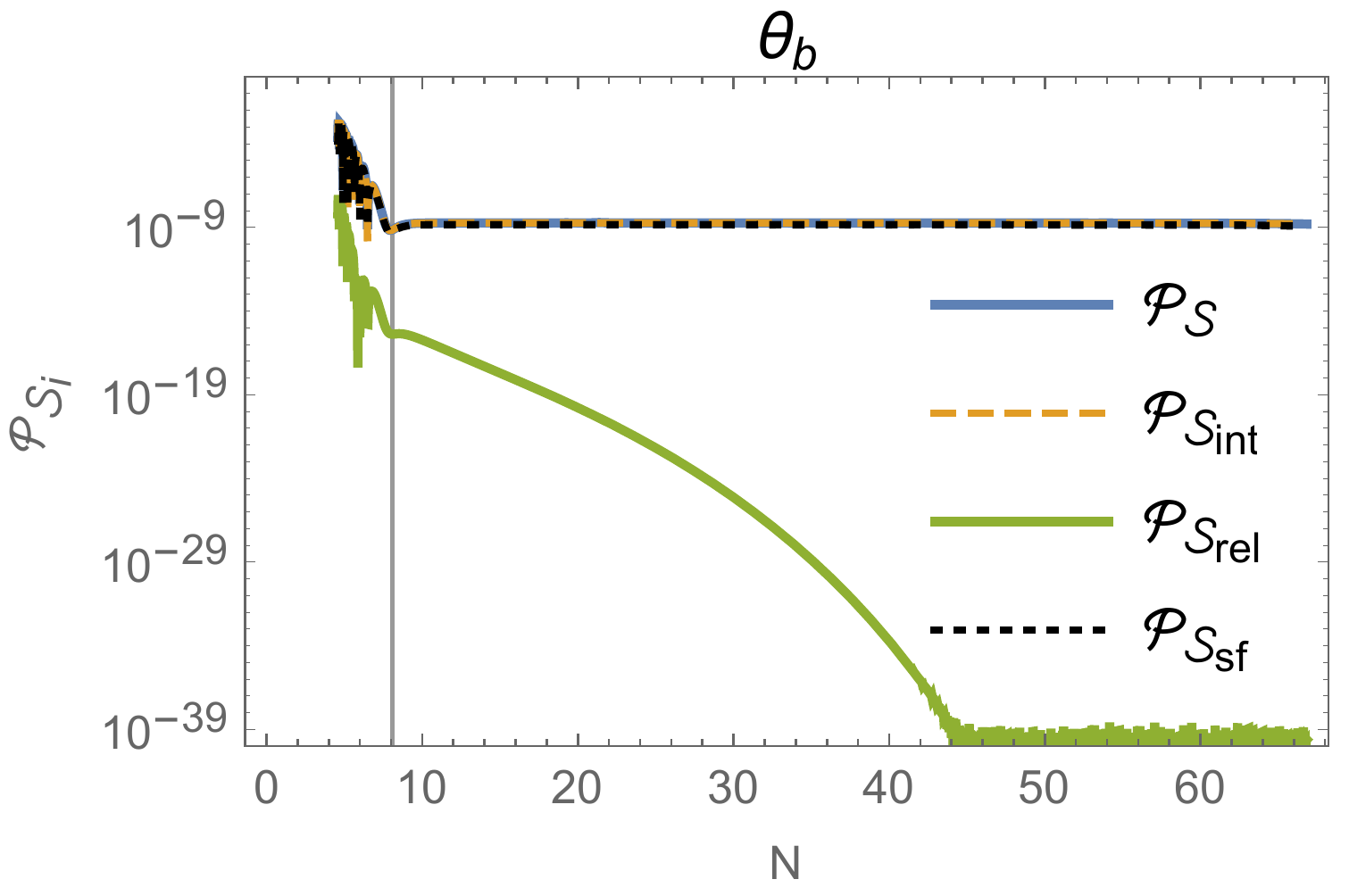}
\includegraphics[scale = 0.5]{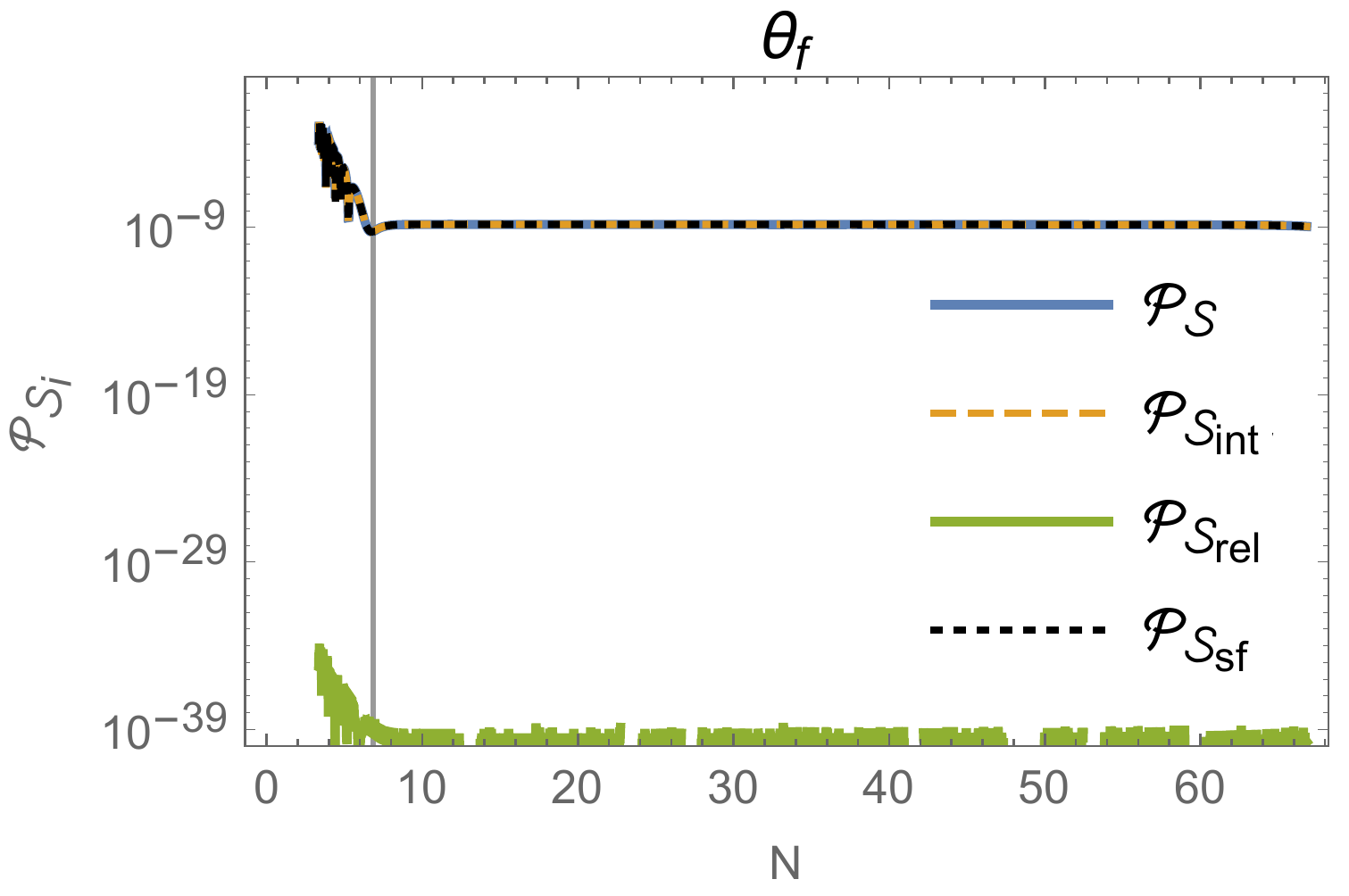}
\caption{Time evolution of different entropy perturbations for modes exiting the horizon $60$ e-foldings before the end of inflation for the 2-field systems $(\phi,\theta_b)$ and $(\phi,\theta_f)$. The choice of initial conditions and microscopic parameters are: $\phi=6.14$, $\phi'=0$, $\theta_i=1$ and $\theta_i'=0.1$ for $i=b,f$, $W_0=25$, $n=4$ for $\theta_b$, and $n=1$ for $\theta_f$ (green line in Fig. \ref{Fig2}). Notice that the results are in practice independent on the axionic initial conditions since $V_b(\phi,\theta_b)$ can reach at most $V_b\sim 10^{-4}\,V_{\rm inf}$ at the beginning of inflation, while $V_f(\phi,\theta_f)$ can go at most up to $V_b\sim 10^{-30}\,V_{\rm inf}$ at the end of inflation. In both cases the only relevant contribution comes from the intrinsic entropy perturbation of the inflaton that coincides with the single field contribution $\mc{S}_{\rm sf}$.}
\label{fig:entropy}
\end{figure}

We have confirmed this analytic result via a numerical analysis (including non-zero axionic potentials) whose results are presented in Fig. \ref{fig:entropy} which shows the super-horizon evolution of different power spectra associated to $\mc{S}$, $\mc{S}_{\rm int}$ and $\mc{S}_{\rm rel}$ given respectively by (\ref{eq:S}), (\ref{eq:Sint}) and (\ref{Srel}). Clearly the contribution coming from the relative entropy perturbation is strongly subdominant for both ultra-light axions. The total entropy perturbation is instead just given by the intrinsic contribution coming from the inflaton that coincides with the single field result.

\section{Conclusions}
\label{Conclusions}

It is fair to say that FI represents one of the best approaches to derive inflation from string theory from both the theoretical and the observational point of view since this class of constructions features controlled moduli stabilisation with an effective approximate shift symmetry, explicit Calabi-Yau embeddings with D-branes, O-planes and chiral matter, and an inflationary potential of Starobinsky-like type which gives so far the best fit to Planck data.

However FI models have been claimed to be plagued by a dangerous geometrical destabilisation effect due to the curvature of the underlying field space \cite{Cicoli:2018ccr}. Starting from the general discussion of entropy perturbation variables performed in \cite{Cicoli:2021yhb}, in this paper we have argued that FI models are actually free from any geometrical destabilisation and the inflationary dynamics is essentially single-field with a negligible production of isocurvature fluctuations. In fact, we have shown that the exponential growth of isocurvature modes associated to perturbations orthogonal to the background trajectory noticed in \cite{Cicoli:2018ccr} is just an unphysical artifact due to the use of an entropy variable which in this case becomes ill-defined due to the anomalous behaviour of the normal unit vector. When studying the evolution during inflation of isocurvature modes associated to the correct physical quantity, the relative entropy perturbation $\mc{S}_{\rm rel}$, we found instead that the corresponding power spectrum decays on super-horizon scales.

We believe that this results holds not just for FI models but more in general also for any inflationary model where the inflaton is kinetically coupled to ultra-light axion-like fields, a situation which can emerge rather naturally in supergravity and string theory effective setups.

\section*{Acknowledgments}

We would like to thank Katy Clough, Evangelos Sfakianakis and Yvette Welling for useful discussions. FM is funded by a UKRI/EPSRC Stephen Hawking fellowship, grant reference
EP/T017279/1 and partially supported by the STFC consolidated grant ST/P000681/1.

\end{document}